\documentclass[12pt]{article}

\usepackage{epsfig,graphics}
\usepackage{amssymb}

\newcommand{\beqn}{\begin{eqnarray}}
\newcommand{\eeqn}{\end{eqnarray}}
\newcommand{\be}{\begin{equation}}
\newcommand{\ee}{\end{equation}}
\newcommand{\ba}{\begin{array}}
\newcommand{\ea}{\end{array}}
\newcommand{\R}{{\rm\bf R}}
\newcommand{\C}{{\rm\bf C}}
\newcommand{\pa}{\partial}

\newcommand{\re}{\ref}
\newcommand{\ci}{\cite}
\newcommand{\la}{\label}
\newcommand{\bfr}{\begin{flushright}}
\newcommand{\efr}{\end{flushright}}
\newcommand{\bfl}{\begin{flushleft}}
\newcommand{\efl}{\end{flushleft}}
\newcommand{\fr}{\frac}

\newcommand{\ti}{\tilde}
\newcommand{\st}{\stackrel}

\newcommand{\ds}{\displaystyle}
\newcommand{\ts}{\textstyle}

\newcommand{\cA}{{\cal A}}

\newcommand{\5}{{\hspace{0.5mm}}}

\newcommand{\we}{\wedge}
\newcommand{\de}{\delta}\newcommand{\De}{\Delta}

\newcommand{\om}{\omega}

\newcommand{\na}{\nabla}

\newcommand{\Br}{|\kern-.25em|\kern-.25em|}
\newcommand{\brr}{{|\kern-.15em|\kern-.15em|\kern-.15em}\,}
\newcommand{\ddd}{\st{.\kern-.07em.\kern-.07em.}}
\def\N{{\rm I\kern-.1567em N}}                              % Doppel-N
\def\R{{\rm I\kern-.1567em R}}                              % Doppel R
\def\C{{\rm C\kern-4.7pt                                    % Doppel C
\vrule height 7.7pt width 0.4pt depth -0.5pt \phantom {.}}}
\def\Z  {{\sf Z\kern-4.5pt Z}}                              % Doppel Z
\renewcommand{\Pr}{{\bf Proof~}}
\begin{document}

\renewcommand{\theequation}{\thesection.\arabic{equation}}
\newtheorem{theorem}{Theorem}[section]
\renewcommand{\thetheorem}{\arabic{section}.\arabic{theorem}}
\newtheorem{definition}[theorem]{Definition}
\newtheorem{deflem}[theorem]{Definition and Lemma}
\newtheorem{lemma}[theorem]{Lemma}
\newtheorem{example}[theorem]{Example}
\newtheorem{remark}[theorem]{Remark}
\newtheorem{remarks}[theorem]{Remarks}
\newtheorem{cor}[theorem]{Corollary}
\newtheorem{pro}[theorem]{Proposition}

\newcommand{\bd}{\begin{definition}}
\newcommand{\ed}{\end{definition}}
\newcommand{\bt}{\begin{theorem}}
\newcommand{\et}{\end{theorem}}
\newcommand{\bqt}{\begin{qtheorem}}
\newcommand{\eqt}{\end{qtheorem}}

\newcommand{\bp}{\begin{pro}}
\newcommand{\ep}{\end{pro}}

\newcommand{\bl}{\begin{lemma}}
\newcommand{\el}{\end{lemma}}
\newcommand{\bc}{\begin{cor}}
\newcommand{\ec}{\end{cor}}

\newcommand{\bex}{\begin{example}}
\newcommand{\eex}{\end{example}}
\newcommand{\bexs}{\begin{examples}}
\newcommand{\eexs}{\end{examples}}

\newcommand{\bexe}{\begin{exercice}}
\newcommand{\eexe}{\end{exercice}}

\newcommand{\br}{\begin{remark} }
\newcommand{\er}{\end{remark}}
\newcommand{\brs}{\begin{remarks}}
\newcommand{\ers}{\end{remarks}}

\newcommand{\pru}{{\bf Proof~~}}

%%%%%%%%%%%%%%%%%%%%%%%%%%%%%%%%%%%%%%%%%%%%%%%%%%
%%%%%%%%%%%%%%%%%%%%%%%%%%%%%%%%%%%%%%%%%%%%%%%%%%

\begin{titlepage}

\phantom{}

\bigskip

\bigskip

\bigskip

\vspace{1.5cm}

\begin{center}
{\Large\bf
On invariants for the Poincar\'e
\medskip

equations  and applications
\\

}

\vspace{2cm}

{\large Valeriy Imaykin} %\footnote{Supported partly by ...}

\medskip

{\it Diagnostic Equipment of Non-Destructive Control, LLC
\\ Proezd Olminskogo 3A,
Moscow, 129164 Russia}\\
email: ivm61@mail.ru

\medskip

\bigskip

{\large Alexander Komech}

\medskip

{\it Faculty of Mathematics of
Vienna University, Oskar-Morgenstern-Platz 1,
1090 Vienna, Austria\\
and IITP RAS, Bolshoy Karetny per. 19-1, 127051 Moscow, Russia
}\\

email: alexander.komech@univie.ac.at

\medskip

\bigskip

{\large Herbert Spohn}

\medskip

{\it Zentrum Mathematik,
TU M\"{u}nchen\\ Boltzmannstr. 3,
Garching, 85747 Germany}\\

email: spohn@ma.tum.de

\end{center}

\vspace{1cm}

%%%%%%%%%%%%%%%%%%%%%%%%%%%%%%%%%%%%%%%%%%%%%%%%%%

\begin{abstract}
We extend the Noether theory of invariants
to the  Poincar\'e equations.
We apply this extension to
the Maxwell-Lorentz
equations coupled to the
Abraham rotating extended electron with the configuration space $SO(3)$.\\


\end{abstract}

\end{titlepage}

%%%%%%%%%%%%%%%%%%%%%%%%%%%%%%%%%%%%%%%%%%%%%%%%%%
%%%%%%%%%%%%%%%%%%%%%%%%%%%%%%%%%%%%%%%%%%%%%%%%%%

\setcounter{equation}{0}

\section{Introduction}

For the
Maxwell-Lorentz
equations with a rotating charged particle (see Eqs. (\re{mls})-(\re{lt}) below)
Hamilton's least action principle is justified in \ci{IKS14}. The main contribution of \ci{IKS14}
is variational
derivation of the {\it Lorentz torque equation} (\re{lt}).
While the equations (\re{mls})-(\re{lf}) follow by standard Euler-Lagrange
arguments, the Lorentz torque equation (\re{lt}) follows
by variational {\it Poincar\'e equations} \ci{AKN,Poin}.

Our main result is a suitable generalization of the  Noether theory of invariants to the
 Poincar\'e equations on the Lie groups.
Moreover, we apply this
generalization to a {\it formal} derivation of conservation laws for the  Maxwell-Lorentz
equations with a rotating charged particle.
We show that the corresponding "Poincar\'e invariants"
coincide with classical known expressions considered in \ci{Kiess}
where their conservation was shown by direct calculation.

We consider solutions for which all
our formal differentiations and integration by parts
hold true.

%%%%%%%%%%%%%%%%%%%%%%%%%%%%%%%%%%%%%%%%%%%%%%%%%%
%%%%%%%%%%%%%%%%%%%%%%%%%%%%%%%%%%%%%%%%%%%%%%%%%%

\setcounter{equation}{0}

\section{Maxwell-Lorentz equations}

The Maxwell fields $E(x,t)$ and
$B(x,t)$ are generated by motion of a rotating charge. External
fields $E^{ext}$
and $B^{ext}$ are generated by the corresponding external
charges and currents.
Let the rotating charge be centered at the position $q$ with the
velocity $\dot q$.
For simplicity we assume that the mass distribution, $m\,\rho(x)$,
and the charge
distribution, $e\,\rho(x)$, are proportional to each other.
Here $m$ is the total
mass, $e$ is the total charge, and we use a system of units such
that $m=1$ and $e=1$.
The coupling function $\rho(x)$ is a sufficiently smooth radially
symmetric function
of fast decay as $|x|\to\infty$,
$$\rho(x)= \rho_r(|x|).\eqno{(C)}
$$

\subsection{Angular velocity}
Let us denote by $\om(t)\in\R^3$ the angular velocity ``in space'' (in
the terminology of \ci{AKN}) of the charge. Namely, let us fix a
``center'' point $O$ of the rigid body. Then the trajectory of each
fixed point of the body is described by
$$
x(t)=q(t)+R(t)(x(0)-q(0)),
$$
where $q(t)$ is the position of $O$ at the time $t$, and $R(t)\in
SO(3)$. Respectively, the velocity reads
\be\la{dotx}
\dot x(t)=\dot q(t)+\dot R(t)(x(0)-q(0))=\dot q(t)+\dot
R(t)R^{-1}(t)(x(t)-q(t))=\dot q(t)+\om(t)\we(x(t)-q(t)),
\ee
where $\om(t)\in\R^3$ corresponds to the skew-symmetric matrix $\dot
R(t)R^{-1}(t)$ by the rule
\be\la{mv}
\dot
R(t)R^{-1}(t)={\cal J}\om(t):=\left(
\ba{ccc}
0 & -\om_3(t) & \om_2(t) \\
\om_3(t) & 0 & -\om_1(t) \\
-\om_2(t) & \om_1(t) & 0
\ea
\right).
\ee
We assume that $x$ and $q$ refer to a certain Euclidean coordinate
system in $\R^3$, and the vector product $\we$ is defined in this
system by standard formulas. The identification (\re{mv}) of a
 skew-symmetric matrix and the corresponding angular velocity
vector is true in any Euclidean coordinate system of the same
orientation as the initial one.

\subsection{Dynamical equations}
Then the system of
Maxwell-Lorentz equations with spin reads, see \ci{Kiess,Sp}
\be\la{mls}
\dot E=\na\we B-(\dot q+\om\we(x-q))\rho(x-q) \qquad (a)\,, \quad \dot B= - \na\we E \qquad (b)\,,
\ee
\be\la{div}
\na\cdot E(x,t)=\rho (x-q(t)) \qquad\qquad\qquad\,\,\,\,\,\,\,\,\,\,\,\,\, (a)\,, \quad \na\cdot B(x,t)= 0 \,\,\,\,\,\,\, (b)\,,
\ee
\be\la{lf} \ddot q=\int\, [E+E^{ext}+(\dot
q+\om\we(x-q))\we(B+B^{ext})]\rho(x-q) \5 dx,
\ee
\be\la{lt} I\,\dot \om= \int \,
(x-q)\we[E+E^{ext}+(\dot q+\om\we(x-q))\we(B+B^{ext})]\rho(x-q)\5 dx,
\ee
where $I$ is the moment of
inertia defined by \be\la{ib} I=\fr 23 \int \, x^2\rho(x)\5 dx.
\ee Here the
equations (\re{mls}) are Maxwell equations with the corresponding
charge density and
current, equations (\re{div}) are constraints. The back reaction
of the field onto
the particle is given through the Lorentz force equation (\re{lf}),
and the Lorentz
torque equation (\re{lt}) deals with rotational degrees of freedom.

%%%%%%%%%%%%%%%%%%%%%%%%%%%%%%%%%%%%%%%%%%%%%%%%%%%%%%%%%%%%%%%%

\subsection{The variational Hamilton principle}

Let us introduce
{\it electromagnetic potentials} ${\cal A}=(A_0,A)$,
${\cal A}^{ext}=(A_0^{ext},A^{ext})$:
\be\la{pot} B=\na\we
A,\,\,\,E=-\na A_0-\dot A. \ee
\be\la{potext} B^{ext}=\na\we
A^{ext},\,\,\,E^{ext}=-\na A_0^{ext}-\dot A^{ext}. \ee

Next we define the Lagrangian
\beqn
\!\!\!\!\!\!&\!\!\!\!\!\!&\!\!\!\!\!\!L({\cal A},q,R,  \dot{\cal A}, \dot q,\dot R)
=\ds\fr{1}{2}\int\left(E^2(x)-B^2(x)\right)\5 dx+\fr{1}{2}\dot
q^2+\fr12 I\om^2
\nonumber\\
\!\!\!\!\!\!&\!\!\!\!\!\!&\!\!\!\!\!\!-\ds\int[A_0(x)+A_0^{ext}(x)]\rho(x-q)\5 dx
+ \ds\int(\dot q+\om\we(x-q))\cdot
[A(x)+A^{ext}(x)]\rho(x-q)\5 dx,\la{Lagrom}
\eeqn
where $E(x)$ and  $B(x)$ are expressed in terms of $\cA(x)$ and $\dot\cA(x)$
according to (\re{pot}),
and $\om={\cal J}^{-1}\dot R R^{-1}$
by (\re{mv}).

This Lagrangian functional depends on $R$ only trough $\om$
due to the spherical symmetry of the charge and mass distributions (C).
Respectively, the dynamical equations  (\re{mls})--(\re{lt})  involve $R$ only through $\om$ as well.
On the other hand, in the case
of non-radial densities the Lagrangian   and the equations
involve $R$ explicitly, and  the moment of
inertia $I$ becomes a matrix with $x\otimes x$ instead of $x^2$ in (\re{ib}).

The corresponding action
functional has the form
\be\la{act}
S=
S(\cA,q,R):=
\int_{t_1}^{t_2} L({\cal A}(t),q(t),R(t),
\dot{\cal A}(t),\dot q(t),\dot R(t))\,dt
\ee
Then the
Hamilton's least action principle reads
\be\la{lap}
\de S(\cA,q,R)=0,
\ee where the
variation is taken over ${\cal A}(t),q(t)$, $R(t)$
with the boundary conditions
\be\la{varclassbound}(\de{\cal A},\de q,\de R)
\vert_{t=t_1}=(\de{\cal A},\de q,\de R)\vert_{t=t_2}=0.
\ee
{\bf Regular solutions and external potential.}
Everywhere below we consider
{\it regular solutions} to the system (\re{mls})--(\re{lt}).
This means that $q\in C^2(\R,\R^3)$, $\om\in C^1(\R,\R^3)$, and all the involved functions
and fields/potentials are
sufficiently smooth and have
(with all the necessary derivatives) a sufficient decay
as $|x|\to\infty$ so that the
partial integrations below are allowed.

\medskip

In \ci[Theorem 2.1]{IKS14} we have shown that for regular solutions,
the Maxwell-Lorentz system  (\re{mls})--(\re{lt})
is equivalent to the
least action principle (\re{lap})--(\re{varclassbound})
In detail, consider the variational equations
\be\la{lapp} \fr{\de S}{\de
{\cal A}}=0\,\,\,\quad(a),\,\,\,\,\,\,\,\,\,\qquad\fr{\de S}{\de q}=0
\,\,\,\quad(b),\,\,\,\,\,\,\,\,\,\qquad\fr{\de S}{\de R}=0
\,\,\,\quad(c). \ee
Then (\re{lapp}), (a), (b) are equivalent respectively to the standard Euler-Lagrange equations
\be\la{ELA}
\fr{d}{dt}\fr{\de L}{\de\dot{\cal A}}=L_{\cal A}\,\,\,\,\,\,\,(a)\,\,\,\,\,\,\,\,\,\,\,\,\,\,\,\,\,\,\,\, \fr{d}{dt}L_{\dot q}=L_q\,\,\,\,\,\,\,\,\,\,(b)
\ee
for the Lagrangian (\re{Lagrom}).
Further, the equation (\re{ELA}), (a)
is equivalent to the Maxwell equations (\re{mls})
with the constraints (\re{div}), and the equation (\re{ELA}), (b) is equivalent to the
Lorentz force equation (\re{lf}).

\smallskip

Note that the equations (\re{lapp}), (a), (b)
are equivalent to standard Euler-Lagrange equations (\re{ELA})
because the variables ${\cal A}$, $\dot{\cal A}$, $q$, and $\dot q$ vary in the corresponding linear spaces. So, we will call these variables the ``Lagrange variables''.

\smallskip

On the other hand, $R\in SO(3)$, and respectively,
the variational equation (\re{lapp}) (c) cannot be transformed to a
Euler-Lagrange equation since
 $SO(3)$ is not a linear space.
We have shown in \ci[Theorem 2.1]{IKS14} that (\re{lapp}) (c)
is equivalent to the
Lorentz torque equation (\re{lt}) using the variational
{\it Poincar\'e equations} with the Lagrangian $L$ expressed in
 suitable
 coordinates on $TSO(3)$.

 \smallskip

In detail, consider an orthonormal basis $\{e_k\}$ with the right orientation in $\R^3$. Then \be\la{basis}
e_1\we e_2=e_3,\,\,\,e_2\we e_3=e_1,\,\,\,e_3\we e_1=e_2. \ee Let us express the
angular velocity in $\{e_k\}$: $\om(t)=\sum\om_k(t)e_k$. The algebra $so(3)$ of
skew-symmetric $3\times3$ matrices with the matrix commutator is isomorphic to the
algebra $\R^3$ with the vector product, through the isomorphism ${\cal J}$ of (\re{mv}):
\be\la{isomv} \left( \ba{ccc}
0 & -\om_3 & \om_2 \\
\om_3 & 0 & -\om_1 \\
-\om_2 & \om_1 & 0
\ea
\right)={\cal J}(\om_1,\om_2,\om_3).
\ee

Namely, let $A,B\in so(3)$, $a,b\in\R^3$, and
$A={\cal J}a$, $B={\cal J}b$.
Then

\be\la{ABBA}
AB-BA={\cal J}(a\we b).
\ee
Further, $\dot RR^{-1}\in T_E SO(3)$ is the
tangent vector $\dot R$ of $SO(3)$ at the point $R$ translated
to the unit $E$ of $SO(3)$ by the right
translation $R^{-1}$.
By the linear isomorphism (\re{isomv}),
\be\la{dRR}
\dot RR^{-1}=\sum\om_k\ti
e_k,~~~~~~~~~ \ti e_k:={\cal J}e_k.
\ee
Then
\be\la{dR} \dot R=\dot RR^{-1}R=\sum\om_kv_k(R),~~~~~~~~~
v_k(R):=\ti e_k R. \ee
As the result, $\dot R$ has the same coordinates w.r.t. the
vector fields $v_k$ at the point $R$ as $\om$ in the basis $\{e_k\}$. The
fields $v_k(R)$ are right translations of $\ti e_k$ and hence are right-invariant.

In \ci[Lemma 6.1]{IKS14} it is shown that for the vector fields $v_k$ on $SO(3)$ the following commutation
relations hold:
\be\la{cr2}
[v_1,v_2]=-v_3,\,\,\,[v_2,v_3]=-v_1,\,\,\,[v_3,v_1]=-v_2.
\ee
We will identify vector fields with the corresponding opertors of differentiation.
According to the Poincar\'e theory \ci{AKN,Poin}, the equation (\re{lapp}) (c) is equivalent to
the Poincar\'e equations
\be\la{PEE} \fr{d}{dt}
\hat L_{\om_k}(Y(t))=\sum_{ij}c^j_{ik}\om_i(t)\hat L_{\om_j}(Y(t))+v_k\hat L(Y(t)),\,\,\,k=1,2,3,
\ee
where $Y(t):=({\cal A}(t),q(t),  \dot{\cal A}(t), \dot q(t),\om(t))$ and
$\hat L({\cal A},q,  \dot{\cal A}, \dot q,\om)$  is defined as the right hand side of (\re{Lagrom}),
and the constants
 $c^j_{ik}$ arise from
commutation relations
$$
[v_i,v_k](R)=\sum c_{ik}^j v_j(R).
$$
In Appendix A,
we recall the calculation of the Poincar\'e equations (\re{PEE}).
These calculations will be used througough the paper.

Note that the Lagrangian $\hat L$ does not
depend explicitly on $R$, and hence $v_k(\hat L)=0$, $k=1,2,3$.
Then
the corresponding Poincar\'e equations read
\be\la{PEE0}
\fr{d}{dt}\fr{\pa\hat L(\om(t))}{\pa\om_k}
=\sum_{ij}c^j_{ik}\om_i\fr{\pa\hat L(\om(t))}{\pa\om_j},\,\,\,\,k=1,2,3.
\ee
In our case (\re{cr2}) and (\re{const}) imply
$$c^3_{21}=c^1_{32}=c^2_{13}=1,\,\,\,c^2_{31}=c^3_{12}=c^1_{23}=-1,\,\,\,{\rm all\,\, the\,\, rest}\,\,\,c^j_{ik}=0.$$
Thus, we can rewrite (\re{PEE0}) as
\be\la{Pevec}
\fr{d}{dt}\fr{\pa\hat L(\om(t))}{\pa\om}=\om\we\fr{\pa\hat L(\om(t))}{\pa\om},
\ee
where $\fr{\ts\pa\hat L}{\ts\pa\om}$ is the column vector with the components $\fr{\ts\pa\hat L}{\ts\pa\om_k}$, $k=1,2,3$.

\smallskip

We summarize the situation as follows, see \ci{IKS14}. The Lagrangian $\hat L$
depends on two groups of variables: on the ``Lagrangian variables'' ${\cal A}$, $\dot{\cal A}$, $q$, $\dot q$ and on the variables $\om_k$ which we will call the ``Poincar\'e variables''.
The variational
equations (\re{lapp}) (a), (b) imply
the Maxwell-Lorentz equations (\re{mls})--(\re{lf}), while
(\re{lapp}) (c) give the Lorentz torque equations (\re{lt}).

%%%%%%%%%%%%%%%%%%%%%%%%%%%%%%%%%%%%%%%%%%%%%%%%%%%%%%%%%%%%%%%%%%%%%%%%%%%%%%%%%%%%%%%%%%%%%%%%%%%
\setcounter{equation}{0}

\section{Invariants for the  Poincar\'e equations}

When the external fields
possess a symmetry with respect to the Lagrangian variables,
the corresponding
conservation laws are given by the Noether theorem on invariants \ci{Arn%K2013
}. In this section we extend the Noether theory to the Poincar\'e equations.

Let $v_1(g),\dots,v_n(g)$ be  vector fields on an
$n$-dimensional manifold $M$ which are linearly independent at each point $g\in M$.
In particular such vector fields exist for any open region $M\subset\R^n$.
 Then $TM$ is isomorphic to $M\times \R^n$, and any
function $L(g,\dot g)$ on $TM$ can be expressed in the {\it Poincar\'e variables} $g,\om$:
\be\la{pvar}
\hat L(g,\om):=L(g,\dot g),\qquad  \dot g=\sum\om_kv_k(g).
\ee
In \ci{Poin},
 Poincar\'e discovered that
the corresponding Hamilton least action principle is equivalent to the
equations
\be\la{Pe0}
\fr{d}{dt}
\hat L_{\om_k}(g(t),\om(t))=\sum_{ij}c^j_{ik}(g)\om_i\hat L_{\om_j}(g,\om)+v_k(g)\hat L(g,\om),\,\,\,k=1,...,n.
\ee
where the  "structure constants" $c^j_{ik}(g)$ arise from
 commutation relations
$$
[v_i,v_j](g)=\sum c_{ij}^k(g) v_k(g),\qquad g\in M.
$$
see details in Appendix A.
Obviously, (\re{PEE}) is the particular case of (\re{Pe0}).

\smallskip

Here we develop the corresponding
theory of
invariant for the Poincar\'e equations.
Let us start with the energy conservation.

\begin{theorem}\la{eecni}
The "energy"
 \be\la{En}
 E:=\hat L_\om\cdot\om-\hat L=\sum\limits_k\hat L_{\om_k}\om_k-\hat L
 \ee
 is conserved along the paths of the Poincar\'e equations (\re{Pe0}).
\end{theorem}
{\bf Proof }
Let a smooth path $(g(t),\om(t))$ satisfy Poincar\'e equations (\re{Pe0}).
Let us compute
\beqn
\frac{d}{dt}(\hat L_\om\cdot\om-\hat L)&=&\frac{d}{dt}\hat L_\om\cdot\om+\hat L_\om\cdot\dot\om-\hat L_g\cdot\dot g-\hat L_\om\cdot\dot\om
=\sum\limits_k\fr{d}{dt}\hat L_{\om_k}\om_k-\hat L_g\cdot\dot g
\la{dotE}\\
\nonumber
&=&\sum\limits_k(\sum\limits_{ij}c^j_{ik}\om_i\hat L_{\om_j}+v_k(\hat L))\om_k-\hat L_g\cdot\dot g
\eeqn
by (\re{Pe0}). Note that $\hat L_g\cdot\dot g=\hat L_g\cdot\sum\om_kv_k=\sum\om_k\hat L_g\cdot v_k = \sum v_k(\hat L)\om_k$. Thus, we obtain
\be\la{dotE0}
\frac{d}{dt}(\hat L_\om\cdot\om-\hat L)=\sum\limits_k(\sum\limits_{ij}c^j_{ik}\om_i\hat L_{\om_j})\om_k=\sum\limits_j\hat L_{\om_j}\sum\limits_{ik}c^j_{ik}\om_i\om_k=0,
\ee
since $\sum\limits_{ik}c^j_{ik}\om_i\om_k=0$ by skew-symmetry property (\re{sks}) of the coefficients $c^j_{ik}$.\hfill$\Box$

\smallskip

\br
In the Lagrangian case (i.e., when $M$ is a linear space and $\om=\dot g$),
the invariant (\re{En}) coincides with the standard energy functional.
\er

Now let us consider general case
of a one-parametric group of diffeomorphisms $h^s:M\to M$ (in particular, $h^0=Id_M$).

\bd\la{dI}
The {\it Poincar\'e invariant} $I$
and the corresponding "current"  $w=(w_1,...,w_n)$ are defined as
\be\la{PI}
I(g,\om):=\sum\hat L_{\om_k}w_k(g),\qquad
\fr{d h^sg}{ds}\Big|_{s=0}=\sum w_k(g)v_k(g).
\ee
\ed
These definitions generalize the corresponding Noether formulas  \ci{Arn}
to the case of Poincar\'e equations.

Let the Lagrangian $L$ be invariant
with respect to the diffeomorphisms $h^s$, i.e.
\be\la{inv}
L(h^sg,d h^s \dot g)= L(g,\dot g),\qquad (g,\dot g)\in TM, \qquad s\in\R.
\ee
\begin{theorem}\la{ecni}
Let condition (\re{inv}) hold.
Then the function
 $I(g,\om)$ is conserved along the paths of the Poincar\'e equations (\re{Pe0}).
\end{theorem}
\Pr
Let a smooth path $(g(t),\om(t))$ satisfy Poincar\'e equations (\re{Pe0}).
Let us denote $g(s,t):=h^sg(t)$ and write
$$
\dot g(s,t)=
dh^s \dot g(t)=\sum\om_k(s,t)v_k(g(s,t)).
$$
In particular,   $g(0,t)=g(t)$ and $\dot g(t):=\sum\om_k(t)v_k(g(t))$.
By (\re{inv}), the quantity
$$
\hat L(g(s,t),\om(s,t)):=L(g(s,t),\sum_k\om_k(s,t)v_k(g(s,t)))
=L(g(s,t),dh^s\dot g(t))
$$
does not depend on $s$;
here $\om(s,t)=(\om_1(s,t),...,\om_n(s,t))$.
Denote by prime the derivative in $s$, and by dot the derivative in $t$. Then we obtain
\be\la{S}
0=\fr{d}{ds}\hat L(g(s,t),\om(s,t))=\hat L_g\cdot g'+\sum\limits_k\hat L_{\om_k}\om_k'= \hat L_g\cdot g'+\sum\limits_k\hat L_{\om_k}(\sum\limits_{ij}c^k_{ij}\om_iw_j+\dot w_k)=:S
\ee
by the formula (\re{cr}) of Appendix A. First we change the order of summation on the right-hand side:
\be\la{ide}
S=\hat L_g\cdot g'+\sum\limits_k\hat L_{\om_k}\dot w_k+\sum\limits_j(\sum\limits_{ik}c^k_{ij}\om_i\hat L_{\om_k})w_j.
\ee
Next we wish to evaluate
the term $\sum\limits_{ik}c^k_{ij}\om_i\hat L_{\om_k}$
for  $s=0$. Namely, $g(0,t)=g(t)$ together with $\om(0,t)=\om(t)$
satisfy the Poincar\'e equations (\re{Pe0}). Hence, for $s=0$
\beqn\la{hen}
S&=&\hat L_g\cdot g'+\sum\limits_k\hat L_{\om_k}\dot w_k+\sum\limits_j\left(\fr{d}{dt}\hat L_{\om_j}-v_j(\hat L)\right)w_j
\nonumber\\
&=&\sum\limits_k\hat L_{\om_k}\dot w_k+\sum\limits_j\fr{d}{dt}\hat L_{\om_j}\cdot w_j+\hat L_g\cdot g'- \sum\limits_jv_j(\hat L)w_j.
\eeqn
However, the definition of the current $w$ in (\re{PI}) implies that
$$\hat L_g\cdot g'- \sum\limits_jv_j(\hat L)w_j = \sum\limits_j\hat L_g\cdot v_jw_j- \sum\limits_jv_j(\hat L)w_j=0
$$
Therefore, (\re{hen}) gives
\be\la{SS}
S=\sum\limits_k\hat L_{\om_k}\dot w_k+\sum\limits_k\fr{d}{dt}\hat L_{\om_k}\cdot w_k=\fr{d}{dt} (\sum\limits_k\hat L_{\om_k}w_k)=\dot I(t).
\ee
The proof is complete, since $S=0$ by (\re{S}).\hfill$\Box$

\br
 Let  $M=\R^n$ and
the vector fields $v_k=\na_{g_k}$ be the commuting fields of differentiations w.r.t. coordinates $g_k$. Then Poincar'e equations (\re{Pe0}) read as the  Euler-Lagrange equations, and the Poincar\'e invariant
(\re{PI})
 coinsides with the  Noether invariant
$L_{\dot g}\cdot\fr{\ts dh^sg }{\ts ds}\Big|_{s=0}$.
\er

%%%%%%%%%%%%%%%%%%%%%%%%%%%%%%%%%%%%%%%%%%%%%%%%%%%%%%%%%%%%%%%%%%%%%%%%%%%%%%%%%%%

\setcounter{equation}{0}
\section{Invariants for the Lagrange-Poincar\'e equations}

Here we generalize the theory of the previous section
to systems with the configuration space
$Y\times M$, where $Y$ is a Hilbert space either
of finite or infinite dimension, while $M$ is a finite-dimensional manifold
endowed with the vector fields $v_k(g)$
as above. Then $TY\simeq Y\times Y$ and $TM\simeq M\times \R^n$.

 Let $L(X,V,g,\dot g)$ be a differentiable Lagrangian which is defined on $TY\times TM$.
 Let us define
\be\la{LL}
\hat L(X,V, g, \om):=L(X,V, g,\dot g),\qquad \dot g=\sum\om_kv_k(g).
\ee
Let a smooth path $(X(t),V(t)), g(t),\om(t)$ satisfy standard Euler-Lagrange equations w.r.t. the variables $(X,V)$ and Poincar\'e equations w.r.t. the variables $(g,\om)$:
\be\la{LP}
\left\{
\ba{rcl}
\ds\fr d{dt}\hat L_V&=&L_X,\\
\\
\ds\fr d{dt} \hat L_{\om_k}&=&\ds\sum_{ij}c^j_{ik}(g)\om_i  \hat L_{\om_j} +v_k(g)\hat L,\,\,\,k=1,...,n.
\ea\right|
\ee

\bt\la{lpe} Let (\re{LP}) hold. Then the energy
\be\la{ELP}
E:=\hat L_V\cdot V+\hat L_\om\cdot\om-\hat L
\ee
is conserved along the path.

\et
\pru
Differentiating formally, we get
$$
\fr{d}{dt}\Big(\hat L_V\cdot V+\hat L_\om\cdot\om-\hat L\Big)=
$$
$$
=\left(\fr{d}{dt}\hat L_V(X,V)
\cdot V-\hat L_X\cdot\dot X\right)
+\Big(\fr{d}{dt}\hat L_\om(g,\om)
\cdot\om-\hat L_g\cdot\dot g\Big)=0.
$$
Indeed, here
the first bracket of the last line
vanishes by the first equation of (\re{LP}).  The second bracket vanishes
by the second equation  of (\re{LP}) that follows by the calculations
 (\re{dotE})--(\re{dotE0}).\hfill $\Box$
\medskip

 Further, consider a  one-parametric group of diffeomorphisms
 \be\la{h12}
 h^s: (X, g)\mapsto (h^s_1(X),h^s_2(g)).
 \ee
 Let us suppose that the Lagrangian functional is $h^s$-invariant, i.e.
 \be\la{Lh12}
 L(h^s_1X,dh^s_1V, h^s_2g,dh^s_2\dot g)\equiv L(X,V, g,\dot g).
 \ee

\bt\la{lpi}
 Let (\re{LP}), (\re{Lh12}) hold.
Then the sum
\be\la{NP}
\hat L_V\cdot\fr{dh_1^sX }{ds}\Big|_{s=0}+\sum\hat L_{\om_k}w_k(g)
\ee
is conserved along the path.
\et
\pru
 Let $X(s,t):=h^s_1X$, $g(s,t):=h^s_2g$, and let $\om(s,t)$ be defined as above. Then
 $\dot g(s,t)=\sum_k v_k(g(s,t))\om_k(s,t)$, and  formally,
$$
0=\fr{d}{ds}\hat L(X(s,t),\dot X(s,t),g(s,t),\om(s,t))=\hat L_X\cdot X'+\hat L_{\dot X}\cdot\dot X(s,t)' + \hat L_g\cdot g'+\sum\hat L_{\om_k}\om_k'.
$$
At $s=0$, the sum of the first two terms reduces to $\fr{d}{dt}\big(L_{\dot X}\cdot\fr{\ts dh_1^sX }{\ts ds}\Big|_{s=0}\big)$ like in the proof of the standard theorem on Noether invariants \ci{Arn}. The sum of the last two terms transforms to $\fr{d}{dt}\big(\sum\hat L_{\om_k}w_k(g)\big)$ like in the proof of Theorem \re{ecni} (calculations (\re{S})--(\re{SS})). \hfill $\Box$

%%%%%%%%%%%%%%%%%%%%%%%%%%%%%%%%%%%%%%%%%%%%%%%%%%%%%%%%%
%%%%%%%%%%%%%%%%%%%%%%%%%%%%%%%%%%%%%%%%%%%%%%%%%%%%%%%%%

\setcounter{equation}{0}
\section{Conservation laws for Maxwell-Lorentz equations}

We now apply the theory of Noether invariants and Poincare invariants for our system of Maxwell-Lorentz equations with rotating charge.
As above, we denote
\be \la{Lagrom1}
\hat L({\cal A},q,  \dot{\cal A}, \dot q,\om)=
L({\cal A},q,R,  \dot{\cal A}, \dot q,\dot R)
\ee
where $\om=(\om_1,\om_2,\om_3)$
is defined by (\re{dR}),  i.e.,
$\om_k$
are coordinates of
$\dot R$ in the basis $v_1(R),v_2(R),$ $v_3(R)$; recall that $\hat L$ does not depend explicitly on $R$.

%%%%%%%%%%%%%%%%%%%%%%%%%%%%%%%%%%%%%%%%%%%%%%%%%%%%%%%%%%
\subsection{Energy}

Let us note that $L$ does
not depend on $\dot A_0$. By Theorem \re{lpe},  we come
formally
to the following statement:

\begin{cor}\la{MLconsE}
Suppose $A_0^{ext}$ end $A^{ext}$ do not depend  on time. Then the functional
\be\la{Energy}
E({\cal A},q,\dot{\cal A},\dot q,R, \om):=\hat L_{\dot A}\cdot\dot A+\hat L_{\dot q}\cdot\dot q+\hat
L_{\om}\cdot\om-\hat L
\ee
is conserved along the regular solutions of the Maxwell-Lorentz system (\re{mls})--(\re{lf}).
\end{cor}

%%%%%%%%%%%%%%%%%%%%%%%%%%%%%%%%%%%%%%%%%%%%%%%%%%%%%%%%%%%%
\subsection{Momentum}

Let the external field
\be\la{indxj}{\cal A}^{ext}(x)=(A_0^{ext}(x),A^{ext}(x))\,\,\,\,{\rm
do\,\,\, not\,\,\, depend\,\,\, on}\,\,\, x_k
\,\,\,\,{\rm for\,\,\, some}\,\,\,\, k.
\ee
Then the Lagrangian (\re{Lagrom1}) is invariant w.r.t to the one-parametric group of spatial translations
\be\la{gsX}
h^s_k({\cal A}(x),q)=({\cal A}(x-se_k),q+se_k),
\ee
where $e_k\in\R^3$ is the corresponding basis vector.
Since the group acts only on the Lagrange coordinates $X:=({\cal A},q)$, $V:=(\dot{\cal A},\dot q)$, we may
formally  apply the  Noether theory
\ci{Arn,K2013} and obtain

\begin{cor}\la{MLKconsM}
Under the condition (\re{indxj}) the functional
\be\la{mom}
P_k=P_k(X,V,R,\om):=\hat L_V\cdot\fr{dh^s_k X}{ds}\Big|_{s=0}
\ee
is conserved for regular solutions to
the Maxwell-Lorentz system (\re{mls})--(\re{lf}).
\end{cor}

\begin{definition}
$P_k$ is called  $k$-th component of  momentum
of the state $(X,V,R,\om)$.
\end{definition}

%%%%%%%%%%%%%%%%%%%%%%%%%%%%%%%%%%%%%%%%%%%%%%%%%%
\subsection{Angular momentum}

Let the external potential ${\cal A}^{ext}$ be axially
symmetric,
\be\la{Uk}
A_0^{ext}(U_k x)=A_0^{ext}(x),\,\,\,\,A^{ext}(U_k x)=U_kA^{ext}(x),
\ee
where $U_k$ is any rotation around the axis $Ox_k$.

\begin{lemma}\la{44}
Let (\re{Uk}) hold.
Then the Lagrangian (\re{Lagrom}) is invariant w.r.t.
the axial rotations
\be\la{frot} A_0(x)\mapsto
A_0(U_k^{-1}x),\,\,\,A(x)\mapsto U_kA(U_k^{-1}x),\,\,\,\dot A(x)
\mapsto U_k\dot A(U_k^{-1}x),
\ee
\be\la{Rrot} R\mapsto U_kR,\,\,\,\,\dot R\mapsto U_k\dot R,
\ee
\be\la{vrot} q\mapsto U_kq,\,\,\,\dot q\mapsto U_k\dot q.
\ee
\end{lemma}
{\bf Proof } By (\re{pot}) the transforms (\re{frot}) of the
potentials induce the following transforms of the fields:
\be\la{EBrot}
E(x)\mapsto U_kE(U_k^{-1}x),\,\,\,\,B(x)\mapsto U_kB(U_k^{-1}x).
\ee
Further, we have, in operator notations, ${\cal J}\om=\om\we$,
where $\om\we$ is the operator of the vector product by $\om$ in $\R^3$.
Then it is easy to check that ${\cal J}(U_k\om)=U_k{\cal J}(\om)U_k^{-1}$.
Thus, for $\om={\cal J}^{-1}\dot RR^{-1}$ we obtain ${\cal J}\om
=\dot RR^{-1}$ and hence ${\cal J}(U_k\om)=U_k(\dot RR^{-1})U_k^{-1}
=(U_k\dot R)(U_kR)^{-1}$. Finally,
$$
U_k\om={\cal J}^{-1}(U_k\dot R)(U_kR)^{-1}.
$$
This means that the transforms (\re{Rrot}) induce the following
transform of $\om$: \be\la{omrot}\om\mapsto U_k\om.
\ee
Now it is easy to check, in view of axial symmetry of
${\cal A}^{ext}$, the invariance of
$L$ w.r.t. the transforms (\re{EBrot}), (\re{vrot}),
(\re{omrot}),
since $\rho$ is spherically symmetric. \hfill $\Box$
\medskip

Recall that $\ti e_k$ is the image of the basis vector
$e_k$ w.r.t. the isomorphism
(\re{isomv}).
By Lemma \re{44} the Lagrangian $\hat L$ (\re{Lagrom1}) is invariant w.r.t.
the spatial rotations (\re{vrot}), (\re{EBrot}), (\re{omrot}).
In particular, $\hat L$ is invariant
under the transform group $h^s_k=e^{s\ti e_k}\in SO(3)$.

In detail, we have the situation of previous section when $\hat L$ depends on Lagrangian variables $(X;V)=({\cal A},q; \dot{\cal A},\dot q)$ and on Poincar\'e variables $(R,\om)$. The action of this group on the state $(X,R)$ reads

$$
h^s_{k}(X,R)=(\alpha^s_kX,\beta^s_kR):\,\,\,\alpha^s_{k}X=(A_0(e^{-s\ti e_k}x), e^{s\ti e_k}A(e^{-s\ti e_k}x),
e^{s\ti e_k}q);
\,\,\,\,\beta^s_{k}R=e^{s\ti e_k}R.
$$
The currents $w_1^k(R),w_2^k(R),w_3^k(R)$ are defined from
\be\la{curs}
\fr{d\beta^s_{k}R}{ds}\Big|_{s=0}=\sum\limits_{j=1}^3w_j^k(R) v_j(R),
\,\,\,R\in SO(3).
\ee
Hence,
by Theorem \re{lpi} we come to the following statement:

\begin{cor}\la{MLconsAM}
Under the condition (\re{Uk}) the quantity
\be\la{Mk}
M_k=M_k(X,V,R,\om):=
\hat L_V\cdot\frac{\ds d\alpha^s_{k}X}{\ds ds}\Big|_{s=0}+ \sum\limits_{j=1}^3\hat L_{\om_j}w_j^k(R)
\ee
is conserved for regular solutions to
the Maxwell-Lorentz system (\re{mls})--(\re{lf}).
\end{cor}

\begin{definition}
$M_k$ is called $k$-th component of
angular momentum
of the state $(X,V,R,\om)$.
\end{definition}

%%%%%%%%%%%%%%%%%%%%%%%%%%%%%%%%%%%%%%%%%%%%%%%%%%
\setcounter{equation}{0}

\section{Expressions for energy and momenta}

Let us show that the Poincar\'e invariants from previous section
coincide with classical known expressions considered in \ci{Kiess}
(where their conservation was shown by direct calculation).

\begin{pro}\la{conserv}
The invariants for the Maxwell-Lorentz system (\re{mls})--(\re{lf}) read as follows:
\medskip\\
i) The energy  reads
\be\la{energy}
E=\fr12\int(|E(x)|^2+|B(x)|^2)\5 dx+\fr12\dot q^2+\fr12 I\om^2+\int\,A_0^{ext}(x)\rho(x-q)\5 dx.
\ee
ii) The momentum reads
\be\la{PP}
P=\dot q+\int\,E(x)\we B(x)\5 dx+\int\,A^{ext}(x)\rho(x-q)\5 dx.
\ee
iii) The  angular momentum reads
\be\la{MM}
M=q\we\dot q+I\om+\int\,x\we E(x)\we B(x)\5 dx+\int x\we A^{ext}(x)\rho(x-q)\5 dx.
\ee
\end{pro}

\noindent{\bf Proof } i) By (\re{Lagrom1}) and (\re{Lagrom}),
one has
$$
\hat L_{\dot A}\cdot\dot A=-\int\,E\cdot\dot A\5 dx,\,\,\,
\hat L_{\dot q}\cdot\dot q=\dot q^2+\int\,\dot q\cdot(A+A^{ext})\rho(x-q)\5 dx,
$$
and
$$
\hat L_{\om}\cdot\om=I\om^2+\int\,(\om\we(x-q))\cdot(A+A^{ext})\rho(x-q)\5 dx.
$$
Then
\beqn
E&=&\hat L_{\dot A}\cdot\dot A+\hat L_{\dot q}\cdot\dot q+\hat
L_{\om}\cdot\om-\hat L=\fr12\dot q^2+\fr12 I\om^2+\fr12\int(|B|^2-|E|^2)\5 dx
\nonumber\\
\la{en1}
&&+
\int(-E\cdot\dot A+A_0\rho(x-q)\,dx+\int\, A_0^{ext}\rho(x-q))\5 dx.
\eeqn
Since
\beqn
\int(-E\cdot\dot A+A_0\rho(x-q))\5 dx&=&\int\,(-E\cdot\dot A+A_0\cdot\na E)\5 dx
\nonumber\\
&=&-\int\,E(\dot A-\na A_0)\,dx=
\int\,E^2\5 dx,
\eeqn
formula (\re{en1}) reads (\re{energy}).

\bigskip

\noindent ii) Let us compute $P_j$.
Formula (\re{gsX}) implies
$$
\fr{dh^s_j(X)}{ds}\vert_{s=0}=
-(e_j\cdot\na A(x),\,\,\,e_j).
$$
%Since $L$ does not depend on $\dot A_0$, and the map $g_s^1$ leaves $\om$ unchanged,
Then
\beqn
P_j&=&
L_V\cdot\fr{dh^s_j(X)}{ds}\vert_{s=0}=-L_{\dot A}\cdot(e_j
\cdot\na)A+L_{\dot q}\cdot e_j
\nonumber\\
&=&-\int(\na A_0+\dot A)\cdot(e_j\cdot\na)A\5 dx+\dot q\cdot e_j+\int\,e_j
\cdot A\rho(x-q)\5 dx+\int A^{ext}_j\rho(x-q)\5 dx
\nonumber\\
&=&\dot q_j+\int\,A_j\rho(x-q)\5 dx-\int(\na A_0+\dot A)\cdot\pa_j A\5 dx+\int A^{ext}_j\rho(x-q)\5 dx.
\eeqn
By partial integration
\beqn
\int\,A_j(x)\rho(x-q)\5 dx&=&\int\,A_j(\na\cdot E)\5 dx=
\int\,A_j\na\cdot(-\na A_0-\dot A)\5 dx
\nonumber\\
\nonumber\\
&=&\int\,A_j(-\De A_0-\na\dot A)\5 dx=
\int(\na A_0\cdot\na A_j+(\dot A\cdot\na)A_j)\5 dx.
\nonumber
\eeqn
Hence,
\be\la{PPjj}
P_j=\dot q_j+\int(\na A_0\cdot\na A_j+(\dot A\cdot\na)A_j)\5 dx -
\int(\na A_0\cdot\pa_j A+\dot A\cdot\pa_j A)\5 dx + \int A^{ext}_j\rho(x-q)\5 dx.
\ee
On the other hand, the j-th component of the RHS of (\re{PP}) equals
$$
\dot q_j+\int(E\we B)_j\5 dx+  \int A^{ext}_j\rho(x-q)\5 dx.
$$
Insert $E=-\dot A-\na A_0$, $B=\na\we A$ and obtain
$$
\dot q_j+\int A^{ext}_j\rho(x-q)\5 dx+ \int\left((\dot A\cdot\na)A_j-
\dot A\cdot\pa_j A+\na A_0\cdot\na A_j-\na A_0\cdot\pa_j A\right)\5 dx
$$
which coincides with (\re{PPjj}).

\bigskip

\noindent iii) For concreteness let us compute $M_1$.
Then
$$
\alpha^s_{1}(X)=(A_0(e^{-s\ti e_1}x),e^{s\ti e_1}A(e^{-s\ti e_1}x),
e^{s\ti e_1}q).
$$
One has

$$\fr{d\alpha^s_1X}{ds}\Big|_{s=0}  =
(-\ti e_1e^{-s\ti e_1}x\cdot\na)A_0(e^{-s\ti e_1}x),
\ti e_1e^{s\ti e_1}A(e^{-s\ti
e_1}x)
  +  e^{s\ti e_1}(-\ti e_1e^{-s\ti e_1}x\cdot\na)A(e^{-s\ti e_1}x),
\ti e_1e^{s\ti
e_1}q)\vert_{s=0}$$
$$=  (\ti e_1A_0(x),\ti e_1A(x)-(\ti e_1x\cdot\na)A(x),\ti e_1q).
$$
Further,
$$
\fr{d\beta^s_{1}R}{ds}\Big|_{s=0}=\fr{de^{s\ti e_1}R }{ds}\Big|_{s=0} = \ti e_1R = v_1(R)
$$
by definition (\re{dR}) of the fields $v_k(R)$. Hence, for the currents $w_j^1$ of (\re{curs}) we have $w_1^1=1$, $w_2^1=w_3^1=0$.
Then, since
$\hat L$ does not
depend on $\dot A_0$,
\beqn
M_1
&=&\hat L_{\dot A}\cdot(\ti e_1A(x)-(\ti
e_1x\cdot\na)A(x))+\hat L_{\dot q}\cdot(\ti e_1q)+\hat L_{\om_1}
\nonumber\\
&=&\int\,\left(\dot A\cdot(\ti e_1A(x)-(\ti e_1x\cdot\na)A(x))
+\na A_0\cdot(\ti
e_1A(x)-(\ti e_1x\cdot\na)A(x))\right)\5 dx
\nonumber\\
&&+\dot q\cdot(\ti e_1q)+\int(\ti e_1q)\cdot (A+A^{ext})\rho(x-q)\5 dx+I\om\cdot
e_1+\int(e_1\we(x-q))\cdot[A+A^{ext}]\rho(x-q)\5 dx
\nonumber\\
&=&(q\we\dot q)_1+I\om_1+\int(x_2 A^{ext}_3-x_3 A^{ext}_2)\rho(x-q)\5 dx
\nonumber\\
\la{second}
&&+\int(x_2A_3-x_3A_2)\rho(x-q)\5 dx+\int(\dot A+\na A_0)\cdot((0,-A_3,A_2)
+(x_3\pa_2-x_2\pa_3)A)\5 dx.
\eeqn
We have to prove that this expression
equals to the first component of the RHS of (\re{MM}).
It suffices
to prove that
the last line (\re{second})
equals to the first component of $\ds \int\,x\we(E\we B)\5 dx$.
Indeed, $\rho(x-q)=\na\cdot E=\na\cdot(-\na A_0-\dot A)$,
hence
\be\la{first}
\!\!\!
\int\!(x_2A_3\!-\!x_3A_2)\rho(x\!-\!q)\5 dx\!=\!
\int\!(x_2A_3\!-\!x_3A_2)(\!-\!\na\dot A\!-\!\na^2A_0)\5 dx\!=\!\int\!\na(x_2A_3\!-\!x_3A_2)(\dot A\!+\!\na A_0)\5 dx.
\ee
Then the line (\re{second}) transforms to
$$
\int\left(\pa_1(x_2A_3-x_3A_2)(\dot A_1+\pa_1A_0)+x_2\pa_2A_3(\dot
A_2+\pa_2A_0)-x_3\pa_3A_2(\dot A_3+\pa_3A_0)\right)\5 dx~~~~~~
$$
\be\la{summa}
+\int\left((x_3\pa_2-x_2\pa_3)A_1(\dot A_1+\pa_1A_0)-x_2\pa_3A_2(\dot
A_2+\pa_2A_0)+ x_3\pa_2A_3(\dot A_3+\pa_3A_0)\right)\5 dx. \ee

On the other hand,
substitute $E=-\dot A-\na A_0$, $B=\na\we A$ and obtain that the first
component of
$\ds\int x\we(E\we B)\5 dx$ equals

$$
~~\int x_2((\pa_1A_3-\pa_3A_1)(\dot A_1+\pa_1A_0)+(\pa_2A_3-\pa_3A_2)(\dot
A_2+\pa_2A_0))\5 dx
$$
$$
-\int x_3((\pa_3A_2-\pa_2A_3)(\dot A_3+\pa_3A_0)+(\pa_1A_2-\pa_2A_1)(\dot
A_1+\pa_1A_0))\5 dx
$$
which coincides with (\re{summa}). The proof is complete.\hfill $\Box$

%%%%%%%%%%%%%%%%%%%%%%%%%%%%%%%%%%%%%%%%%%%%%%%%%%
%%%%%%%%%%%%%%%%%%%%%%%%%%%%%%%%%%%%%%%%%%%%%%%%%%

\appendix

%%%%%%%%%%%%%%%%%%%%%%%%%%%%%%%%%%%%%%%%%%%%%%%%%%

\setcounter{equation}{0}

\section{Poincar\'e equations}

Poincar\'e suggested the form of
the Hamilton least action principle for Lagrangian systems on
manifolds \ci{Poin}. We present
the derivation of the Poincar\'e equations \ci{AKN} since we use
some of intermediate calculations.

Let $v_1,\dots,v_n$ be vector fields on a $n$-dimensional manifold $M$
which are
linearly independent at every point $g\in M$. Then the commutation
relations hold,
$$[v_i,v_j](g)=\sum c_{ij}^k(g)v_k(g),~~~~~~~g\in M
$$
where the commutator $[v_i,v_j]$ is defined by
$$
[v_i,v_j](f):=v_i(v_j(f))-v_j(v_i(f)),
$$
and $v(f)$ is the derivative of a smooth function $f$ on $M$ w.r.t.
the vector field $v$. Note that by the skew-symmetry property of the commutators one has
\be\la{sks}
c_{ij}^k(g)=-c_{ji}^k(g),\,\,\,\forall\,\,k=1,...,n.
\ee
If $g(t)$ is a smooth path in $M$
and $f$ is a smooth function on $M$, one has
$\dot g(t)=\sum\om_i(t)v_i(g(t))$ and
$$
\fr{d}{dt}f(g(t))=f'(g(t))\cdot\dot g=
f'(g(t))\cdot\sum\om_i(t)v_i(g(t))
=\sum
v_i(f)\om_i(t).
$$
Now
consider a variation $g(s,t)$ of the path $g(t)$. Then
similarly,
$$
\pa_s f(g(s,t))=\sum_jv_j(f)w_j(s,t),
$$
where $w_j(s,t)$ are coordinates of $\fr{\pa g}{\pa s}(s,t)
\in T_{g(s,t)}M$.
Hence
$$
\pa_s\pa_t
f(g(s,t))=\sum_i\sum_jv_j(v_i(f))w_j\om_i
+\sum_iv_i(f)\om_i',
$$
$$
\pa_t\pa_s
f(g(s,t))=\sum_j\sum_iv_i(v_j(f))w_j\om_i
+\sum_jv_j(f)\dot w_j,
$$
where the prime respectively the dot stand for the differentiation in $s$
respectively in $t$. However, the differentiations in $t$ and $s$ commute, hence we
obtain by subtraction
$$
\sum_k v_k(f)\om_k'=\sum_k\sum_{ij}c^k_{ij}\om_iw_jv_k(f)
+\sum_k v_k(f)\dot w_k.
$$
Since $f$ is an arbitrary smooth function, we come to the
 relations
\be\la{cr} \om_k'(s,t)=\sum_{ij}c^k_{ij}\om_iw_j+\dot w_k.
\ee

Further, let us consider a Lagrangian function $L(g,\dot g)$ on $TM$.
Then $L(g,\dot g)$ can be expressed in the variables $\om$: $L(g,\dot g)=\hat L(g,\om)$.
Let us
compute the variation of the corresponding action functional
taking (\re{cr})
into
account:
$$
\fr{d}{ds}\int_{t_1}^{t_2}\hat L(g(s,t),\om(s,t))dt=
\int_{t_1}^{t_2}\left(\sum_k\fr{\pa\hat L}{\pa\om_k}\om_k'+ \na_g\hat L\cdot
g'\right)dt=
$$
$$
\int_{t_1}^{t_2}\left[\sum_k\fr{\pa\hat L}{\pa\om_k}(\dot
w_k+\sum_{ij}c^k_{ij}\om_iw_j) +\na_g\hat L\cdot\sum_kw_kv_k\right]dt=
$$
$$
\sum_k\fr{\pa\hat L}{\pa\om_k}w_k\Big|_{t_1}^{t_2}+\int_{t_1}^{t_2}
\sum_k\left[-\fr{d}{dt}\fr{\pa\hat L}{\pa\om_k}
+\sum_{ij}c^j_{ik}\om_i\fr{\pa\hat
L}{\pa\om_j}+v_k(\hat L)\right]w_k\,dt.
$$

The variation should be zero by the Hamilton least action principle, under the
boundary value conditions
\be\la{star}
g(s,t_1)=g_1,\,\,\,g(s,t_2)=g_2.
\ee Since
$w_k(t_1)=w_k(t_2)=0$ by (\re{star}), we obtain the following
{\it Poincar\'e
equations}:
\be\la{Pe}
\fr{d}{dt}\fr{\pa\hat
L}{\pa\om_k}=\sum_{ij}c^j_{ik}\om_i\fr{\pa\hat L}{\pa\om_j}+v_k(\hat L).
\ee {\bf
Remarks } 1. If $g$ is expressed in a local map as
$(g_1,...,g_n)\in\R^n$, and $v_k=\pa_{g_k}$, then (\re{Pe}) reduce
to the standard Euler-Lagrange equations.

\smallskip

\noindent 2. If a Lagrangian $L$ does not depend on $g$,
then $\hat L=\hat L(\om)$ and one has
\be\la{Lom} v_k(\hat L)=0. \ee Indeed, $v_k(\hat L)=\na_g\hat L\cdot v_k(g)=0$.

\smallskip

\noindent 3. Suppose $M=G$ is a Lie group, and  let $v_k$, $k=1, ...,\,n$ be
independent either left-invariant or right-invariant vector fields on $G$.
Then $c^k_{ij}(g)$ are constant:

\be\la{const} c^k_{ij}(g)\equiv c^k_{ij},~~~~~~~~\,g\in G.
\ee

\bigskip

{\bf Acknowledgements } The research of A. Komech was
supported partly by Austrian Science Fund (FWF): P28152-N35.
The research of V. Imaykin and A. Komech were supported partly by the grant of RFBR: 16-01-00100. 
The authors thank Professor M.~Kiessling for useful
discussions and remarks.

%%%%%%%%%%%%%%%%%%%%%%%%%%%%%%%%%%%%%%%%%%%%%%%%%%
%%%%%%%%%%%%%%%%%%%%%%%%%%%%%%%%%%%%%%%%%%%%%%%%%%

%%%%%%%%%%%%%%%%%%%%%%%%%%%%%%%%%%%%%%%%%%%%%%%%%%
%%%%%%%%%%%%%%%%%%%%%%%%%%%%%%%%%%%%%%%%%%%%%%%%%%
\end{document}